\documentstyle[12pt,epsf]{article}

\def\Journal#1#2#3#4{{#1} {\bf #2}, #3 (#4)}

\def\ffrac#1#2{\textstyle{#1\over#2}\displaystyle}
\def\rg{renormalisation\ group}
\def\ad{{a^{\dag}}}
\def\as{{a^*}}
\def\ab{{\bar a}}
\def\cd{{c^{\dag}}}

\def\be{\begin{equation}}
\def\ee{\end{equation}}
\def\bea{\begin{eqnarray}}
\def\eea{\end{eqnarray}}
\begin{document}

\title{Renormalisation Group Approach To Reaction-Diffusion Problems} 

\author{John Cardy\\
{}\\
All Souls College and\\
Department of Physics, Theoretical Physics, 1 Keble Road,\\
Oxford OX1 3NP, England}
\maketitle

\begin{abstract}
We describe how the methods of quantum field theory and the
renormalisation group may be applied to classical stochastic particle
systems which appear in non-equilibrium statistical mechanics. An
emphasis is placed on the similarities and differences between these
methods and more conventional applications of quantum field theory.
Some simple applications are discussed.
\end{abstract}

\section{Introduction}
Reaction-diffusion problems are simple examples of non-equilibrium
statistical systems. While it has long been recognised that the methods of
quantum field theory extend far beyond their original domain of
application in elementary particle physics, to, for example, many body
quantum systems in condensed matter physics, and to the theory of both
the statics and dynamics of critical behaviour in equilibrium
statistical mechanics, it is less well appreciated that they also provide a
powerful tool for analysing classical statistical systems far from
equilibrium. This application is certainly not 
new,\cite{Doi,Peliti,Grassberger} but it is only
relatively recently that the full formalism of the \rg\ has been brought
to bear,\cite{Lee,CardyLee,Howard} in a
very similar manner to that employed in equilibrium critical behaviour.

Reaction-diffusion problems are examples of classical stochastic
particle systems. These particles are in general labelled by their
species $A,B,\ldots$, which may be thought of as corresponding to
different chemical reactants. Their dynamics consists of two elements:
first, the particles diffuse according to some kind of Brownian motion,
with diffusion constants $D_A,D_B,\ldots$. In a simulation, this might
be described by random walks on a lattice. Second, they undergo
reactions, for example $A+B\to C$, at prescribed rates $\lambda_{AB}^C$,
whenever they meet. It is important that these processes are \em
diffusion-limited\em, that is, the particles have to diffuse around
before they find each other and react. In a chemical system this would
mean that the reactants should not be stirred. In practice this may be
realised by allowing the reactants to diffuse in a gel, or on a
substrate. It is important to realise that the these reactions are not
usually reversible, and indeed the most interesting cases occur when
they are completely irreversible. 

The potential applications of these ideas to systems in chemistry,
biology and physics are limited only by the imagination of the reader,
and this is not the place to discuss them. Rather, we shall focus on
some general features as illustrated by some rather simple examples. 
In general, from some random initial state, these systems will evolve in
time towards a steady state described by some stationary probability
distribution. Since the dynamics does not satisfy detailed balance, this
is not in general a Gibbs measure. In some very simple cases, the steady
state is trivial. For example, in the simplest system of all, with a
single species of particle $A$ undergoing the pure annihilation reaction
$A+A\to\,$inert (the inert particles not influencing the reaction
further), the steady state has no particles, and hence no fluctuations
since it is a classical vacuum. But it turns out that the \em approach
\em to this steady is critical in the sense that it exhibits universal
scaling behaviour, critical exponents, and so on. This may be
understood, and the universal features computed, within the quantum
field theoretic renormalisation group approach to be described. 

A second class of critical phenomena corresponds to a non-equilibrium
phase transition in the steady state, as some parameter of the dynamics
is varied. For example, if to the annihilation reaction described above
the branching process $A\to(m+1)A$ at rate $\sigma$ is added, it turns
out that under some circumstances there can be a transition at some
finite value of $\sigma$ to a non-trivial stationary state with a finite
density of particles.\cite{Grassberger,TT,Jensen}.
The universality class of this transition turns
out to depend on the parity of $m$. As with equilibrium critical
behaviour, the symmetries of a system are seen to play an
important role in determining its universality class.

The layout of this paper is as follows. In the next section we describe
the general formalism and illustrate it with the annihilation reaction
$A+A\to\,$inert. We comment on the similarities and the differences with
ordinary many-body quantum theory, and on the connections between this
approach to non-equilibrium behaviour and others via the Fokker-Planck 
equation and the Langevin equation. In Sec.~\ref{AA} we discuss the
renormalisation group approach to the $A+A\to\,$inert reaction in more
detail, and finally in Sec.~\ref{BARW} we consider the more difficult
problem of when the branching process is added. 
\section{Formalism}\label{formalism}
The dynamics of such a stochastic particle system is described by a master
equation governing the time evolution of the probabilities 
$p(\alpha;t)$ that the system be in a given microstate $\alpha$. For a
system of particles on a lattice, for example, the $\alpha$s might
label the occupation number basis $(n_1,n_2,\ldots)$,
corresponding to having $n_j$ particles at site $j$. The master equation
takes the form
\be\label{me}
dp(\alpha;t)/dt=\sum_\beta R_{\beta\to\alpha}\,p(\beta;t)
-\sum_\beta R_{\alpha\to\beta}\,p(\alpha;t).
\ee
Here $R_{\alpha\to\beta}$ is the rate for transitions from state
$\alpha$ into $\beta$; for a reaction diffusion-problem these are
determined by the diffusion constants and the reaction rates.

Such classical particle problems have two features in common with
relativistic many-body problems which renders a `second-quantised'
formalism particularly useful. First, particle numbers are not, in
general, conserved: they are created and destroyed by the dynamics.
Second, and more important, the master equation and the Schr\"odinger
equation share the properties of being linear and first order in time.
It is therefore not surprising that such a formalism is similarly
successful for these stochastic problems. 

The first step is to construct a Fock space from annihilation and
creation operators satisfying the usual commutation relationships
$[a_i,\ad_j]=\delta_{ij}$, and define the state vector
\be
|\Psi(t)\rangle\equiv\sum_\alpha p(n_1,n_2,\ldots;t)\,\ad_1^{n_1}
\ad_2^{n_2}\ldots|0\rangle.
\ee
Note that this is not normalised in the conventional manner, and what
plays the role of a probability amplitude in quantum mechanics now is a
probability. We have also chosen a bosonic representation, corresponding
to the case when multiple occupancy of the sites is allowed. In
simulations it is often more convenient to restrict the values of the
$n_j$ to 0 or 1, in which case a representation in terms of Pauli
operators is more appropriate. This leads to quantum spin models rather
than immediately to quantum field theories, which, in some
one-dimensional cases, turn out to be integrable. In fact there is a
some very elegant mathematics in this branch of the subject.\cite{Rittenberg}
For example
the quantum group symmetry of certain spin chains appears very naturally
from this perspective. However, from the point of view of
understanding the \rg\ approach and generalising to noninteger
dimensions, the bosonic formulation is more useful. In any case, as long
as we are studying problems where the average particle density is low,
the probability of multiple occupation should be small and there should
be no difference between the physical results of the two approaches.

The statement is now that the master equation (\ref{me}) is completely
equivalent to a Schr\"odinger-like equation for the state vector
\be\label{se}
d|\Psi(t)\rangle/dt=-H|\Psi(t)\rangle,
\ee
where the `hamiltonian' $H$ is simply expressed in terms of the $a$s and
the $\ad$s. For example, for the reaction-diffusion problem
$A+A\to\,$inert, one finds that
\be\label{HAA}
H=D\sum_{(i,j)}(\ad_i-\ad_j)(a_i-a_j)
-\lambda\sum_i(a^2_i-\ad_i^2a_i^2).
\ee
The simple hopping form of the first term (where the sum is over nearest
neighbour pairs $(i,j)$), which corresponds to pure diffusion, is not
surprising, but the second term may require some explanation. To
understand its form, consider the simpler problem of the reaction
at a single site.
If $p(n;t)$ is probability of finding $n$ particles at this site, the
master equation is simply
\be
dp(n;t)/dt=\lambda(n+2)(n+1)p(n+2;t)-\lambda n(n-1)p(n;t),
\ee
where the factors of $(n+2)(n+1)$ reflect the number of ways of choosing
the pair of reacting particles.
Defining $|\Psi\rangle=\sum_np(n;t)\ad^n|0\rangle$ as above, its
equation of motion is
\bea
d|\Psi\rangle/dt&=&\lambda\sum_n\Big((n+2)(n+1)p(n+2)-n(n-1)p(n)\Big)
\ad^n|0\rangle\\
&=&\lambda\sum_n\Big(a^2p(n+2)\ad^{n+2}-\ad^2a^2p(n)\ad^n\Big)|0\rangle\\
&=&\lambda(a^2-\ad^2a^2)\sum_np(n)\ad^n|0\rangle.
\eea
The second term in the reaction part of (\ref{HAA}) therefore
corresponds to the second term in the master equation (\ref{me}), and is
required by the conservation of probability.

From the lattice hamiltonian (\ref{HAA}) we may, if interested in long
wavelength properties, proceed to the formal continuum limit
\be\label{ham}
H=\int\big[D(\nabla\ad)(\nabla a)-\lambda(a^2-\ad^2a^2)\big]d^d\!x,
\ee
and thence to a representation as a path integral over fields
$a(x,t)$ and $\as(x,t)$ with a weight $\exp(-S[a,\as])$, with an action
\be
S\equiv\int\big[a^*\partial_ta+D(\nabla a^*)(\nabla a)-
\lambda(a^2-{a^*}^2a^2)\big]dtd^d\!x.
\ee
\subsection{Differences from quantum mechanics}
There are two immediately apparent differences from ordinary quantum
field theory: first, there is no factor of $i$ in the Schr\"odinger
equation (\ref{se}) -- but this is familiar from euclidean formulations
of conventional quantum theories; second, the hamiltonian is not
hermitian. In many cases it will turn out that, nevertheless, its 
eigenvalues are real. (Complex eigenvalues correspond to oscillating
states which are known to occur in some chemical reactions.) However,
the most important difference is one of interpretation: expectation
values of observables $\cal O$ are not given by 
$\langle\Psi|{\cal O}|\Psi\rangle$, since this would be bilinear, rather
than linear, in the probabilities $p(\alpha;t)$. Instead, for an
observable which is diagonal in the occupation number basis, 
its expectation value is of course
\be
\overline{\cal O}=\sum_{\{n_j\}}{\cal O}(\{n_j\})p(\{n_j\};t) ,
\ee
and it is straightforward to show that this may be expressed as
\be\label{ev}
\overline{\cal O}=\langle0|e^{\sum_ja_j}{\cal O}\,e^{-Ht}|\Psi(0)\rangle,
\ee
since the state $\langle0|e^{\sum_ja_j}$ is a left eigenstate of all the
$\ad_j$, with unit eigenvalue.

Conservation of probability then requires that
$\langle0|e^{\sum_ja_j}H=0$. This is equivalent to the requirement that
$H$ should formally vanish when every $\ad_j$ is set to unity. The
appearance of the state $\langle0|e^{\sum_ja_j}$ may complicate some of the
subsequent calculations, since the interaction part of the hamiltonian
is not normal ordered with respect to it, and therefore the usual
formalism of time-dependent perturbation theory and Wick's theorem do
not immediately apply. This problem may be avoided by first commuting
the factor of $e^{\sum_ja_j}$ through the operators $\cal O$ and $H$ in
(\ref{ev}). This has the effect of shifting $\ad\to1+\ad$, since
$e^a\ad=(1+\ad)e^a$. The factor $e^{\sum_ja_j}$ acting on the initial
state $|\Psi(0)\rangle$ is usually something simple, and the operators
are now normal ordered. 

Note that such a shift is convenient only if we are interested in what
(in the language of particle physics) may be termed `inclusive'
probabilities, for example the expectation value of the local density
$\bar n_j=\overline{\ad_ja_j}$. After the shift, we see that in fact
$\bar n_j=\langle a_j\rangle$, where $\langle\cdot\rangle$ denotes the
usual QFT expectation value. For so-called exclusive quantities, for
example the probability $\overline{\delta_{n_i1}\prod_{j\not=i}
\delta_{n_j0}}$ that there is only one particle in the system, at site
$i$, the factor $e^{\sum_ja_j}$ simply reduces to $a_j$ in the
correlation function, and no shift is necessary.
\subsection{Relation to other formalisms}
Of course, there are several other important ways of formulating
stochastic processes, through either the Langevin equation or its
related Fokker-Planck equation. It is interesting to see how these
emerge in the present formalism for the simple example under
consideration. If we make the shift $\as=1+\ab$ in the
path integral, we find an action
\be\label{Sshift}
S[a,\ab]=\int\big[\ab\partial_ta+D(\nabla\ab)(\nabla a)
+2\lambda\ab a^2+\lambda\ab^2a^2\big]dtd^d\!x.
\ee
The non-linear terms in $\ab$ may be disentangled in terms of a Gaussian
transformation 
\be
e^{-\lambda\ab^2a^2}=\int e^{-\eta\ab}P(\eta)d\eta,
\ee
where $P(\eta)$ is suitable Gaussian distribution. The path integral
over $\ab$ may now be performed, yielding a functional delta function
equivalent to the equation 
\be\label{langevin}
\partial_ta=D\nabla^2a-2\lambda a^2+\eta(x,t).
\ee
Neglecting the last term, this is just the so-called rate equation which
one might write down as a first approximation to the equation of motion
for the density (note that we have argued above that the expectation
value of $a$ is the density.) The rate equation approximation assumes
that the annihilation rate, which is proportional to the probability of
finding two particles at the same point, is simply given by the square
of the density. This approximation clearly neglects the correlations
between the particles, and is on the same footing as the mean field
approximation in equilibrium critical behaviour. In this sense 
(\ref{langevin}) looks very much like a Langevin equation, with a noise
term $\eta$. However, such
equations are usually derived from the master equation through some kind
of approximate coarse-graining, and the exact form of the noise term
is often unclear, especially when the dynamics does not constrain this
through detailed balance. By contrast, the correlations of the noise
here are completely explicit
\be
\langle\eta(x,t)\eta(x',t')\rangle=-\lambda a^2\delta(x-x')\delta(t-t').
\ee
That the noise should depend on $a$ is expected, since there can
be no noise when the density is zero. But the minus sign is surprising.
It implies that the noise $\eta$ is pure imaginary, so that the solution
of (\ref{langevin}) is complex! 

This curious result may be traced to the fact that, although the 
`quantum mechanical' average $\langle a\rangle$ is the mean density
$\bar n$, this is not true of higher moments. For example, the mean
square density $\overline{n^2}$ is given by the average of
$(\ad a)^2=\ad^2a^2+\ad a$. The operators $\ad$ give unity 
acting to the left on the state $\langle0|e^a$, 
so that in fact $\overline{n^2}=\langle a^2\rangle+\langle a\rangle$.
In general, one may show that, if $a$ has a Gaussian distribution, as it
would in the case of pure diffusion, then the density $n$ would have a
pure Poisson distribution. This is to be expected, since a simple random
walk will have Poissonian statistics. The effect of the reactions is to
modify this. In a sense the `noise' $\eta$ in (\ref{langevin})
represents only that part of the physical noise which originates in the
discreteness of the reaction process. Since, however, this cannot truly
be disentangled from the diffusion noise, there is no need for its
correlations to be positive.

The hamiltonian approach we have described above should be distinguished
from another based on the Fokker-Planck equation. The latter begins from
a coarse-grained Langevin description of the problem, and describes the
time evolution of the probability distribution of the solution of this
equation. Like the master equation, it is linear and first order in
time, so may be cast in a hamiltonian formalism (in this case
more usually called the liouvillean.) But the Fokker-Planck equation
describes the diffusion in phase space and cannot easily accommodate
processes where particles are continually being created and destroyed.
It is therefore less useful for these types of problem.
\section{Renormalisation group analysis}\label{AA}
\begin{figure}
\centerline{
\epsfxsize=4.5in
\epsfbox{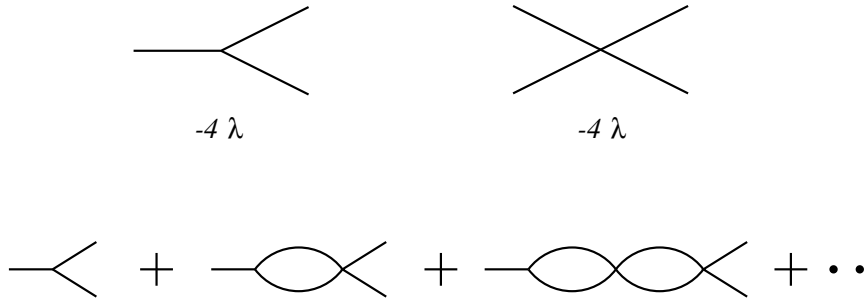}}
\caption{Vertices and renormalisation for $A+A\to\,$inert.}
\label{FR}
\end{figure}

The field theory described by the action (\ref{Sshift}) is extremely
simple. The bare propagator $(-i\omega+Dq^2)^{-1}$
is simply the Green function for the diffusion equation, and it may
be represented by a directed line moving forward in time (conventionally
from right to left.) The vertices are shown in Fig.~\ref{FR}.
Immediately we see that, since the number of particles in a given
intermediate state cannot increase as we move from right to left through
a diagram, there can be no loop corrections to the
propagator, hence no wave function renormalisation for the fields $a$ or
$\ab$.  
The only non-trivial renormalisation is that of the coupling constant
$\lambda$, which may be seen through the loop corrections either to the
vertex $\Gamma^{2,1}$ (shown in Fig.~\ref{FR}) or to $\Gamma^{2,2}$.
(The fact that these renormalise in the same way is a consequence of
probability conservation, which relates the untruncated Green
functions $\langle0|a(t,x)\ad^2(0,0)|0\rangle$ and
$\langle0|a^2(t,x)\ad^2(0,0)|0\rangle$. It is interesting to note that
probability conservation is not expressed through a Noether symmetry in
this formalism.) The diagrams in Fig.~\ref{FR} correspond to a simple
inclusion-exclusion argument for the probability of finding two
particles at the same point given that they have not reacted in the
past. They may be simply summed, with the result that the one-loop
\rg\ beta function is exact:
\be
\beta(g_R)=-\epsilon g_R+bg_R^2,
\ee 
where $g_R$ is the dimensionless renormalised coupling,
$b$ is a positive constant, and $\epsilon=2-d$. 

For $d>2$, then, $g_R$ is irrelevant in the infrared, and the rate
equation (\ref{langevin}) with no noise term is asymptotically valid, while
for $d<2$ it flows to a non-trivial fixed point $g^*=O(\epsilon)$. The
consequences of this may be explored by writing down and solving the
Callan-Symanzik equation in the usual way. Consider, for example, the
mean density $n(t)$, which depends in principle on the initial density
$n_0$ and the rate $\lambda$, expressed through $g_R$ and the
normalisation scale $t_0$. (We work in units where the diffusion
constant $D=1$.) Then
\be
n(t,n_0,g_R,t_0)=(t/t_0)^{-d/2}\,n(t_0,n_0(t/t_0)^{d/2},\tilde
g_R(t/t_0),t_0)
\ee
where the running coupling $\tilde g_R\to g^*$ as $t/t_0\to\infty$.
The simple exponent in the prefactor on the right hand side reflects the
lack of any anomalous dimensions for the density. The right hand side
may now be evaluated as a power series in $g^*$ (the lowest order being
simply the mean field result), which is converted into a power series in
$\epsilon$. Term by term, one may show that the result is in fact
independent of the initial density $n_0$, so that in fact the whole
expression is a universal function of $\epsilon$ only. The result is
$n(t)\sim A/t^{d/2}$ where\cite{Lee}
\be
A={1\over4\pi\epsilon}+ {2\ln8\pi-5\over16\pi} + O(\epsilon).
\ee
Universal forms for correlation functions may be derived in a similar
manner. The result is that the whole probability distribution for
fluctuations becomes universal in the late time regime.
\section{Branching and annihilating random walks}\label{BARW}
A more interesting type of critical behaviour emerges if to the
annihilation reaction $A+A\to\,$inert we add the branching process 
$A\to(m+1)A$, with rate $\sigma$. Here $m$ is a positive integer. 
In the rate equation approximation, the mean density now satisfies
\be
dn/dt=\sigma mn-\lambda n^2,
\ee
which would suggest that, for any $\sigma>0$, the steady state has a
non-zero density of particles (it is `active' in the language of
catalysis.) In fact, simulations suggest that this is only true in
sufficiently high dimensions $d>2$. For $d\leq2$ the fluctuations need
to be taken into account, and this may be done using the formalism
described above.

In one dimension, with $m=2$, this model may be interpreted as a dynamic Ising
model, where the particles play the role of domain walls.\cite{Meynhard}
In that case
the processes of diffusion and annihilation are generated by single spin
flips (these are assumed to occur at zero temperature so that pair
creation of domain walls is suppressed), and the branching process
$A\to3A$ is associated with spin exchange (assumed to occur at infinite
effective temperature). Since this model has two `temperatures', it does
not satisfy detailed balance and the stationary state is not Gibbsian.
For that reason the model may undergo a nontrivial ordering transition,
even in one dimension.

The additional term in the hamiltonian has the form
\be
H_{b}=\sigma\int[\ad^{m+1}a-\ad a]d^d\!x.
\ee
Note that there is a difference depending on the parity of $m$. If it is
even, then the number of particles is locally conserved modulo 2 by both
the annihilation and the branching process, while for $m$ odd the latter
violates this. For $m$ even this is manifested in the formal symmetry of
the hamiltonian under $(a,\ad)\to(-a,-\ad)$. Note that if we make the
shift $\ad\to1+\ab$ in order to develop the perturbative expansion for
`inclusive' processes, this symmetry becomes hidden. If we further make
an expansion of the hamiltonian in powers of $\ab$, and drop higher
orders on the grounds of irrelevance by power counting, the symmetry is
completely lost. This is evidently a dangerous thing to do, since it is
well known that symmetries play a very important role in influencing
universality classes of critical behaviour. Fortunately the
renormalisation group behaviour of the theory should be independent of
which type of correlation functions we choose to study, and therefore
may be computed in the unshifted theory where the symmetry is manifest.
\subsection{Case of even $m$}
The first question to be addressed\cite{Tauber} is whether the branching rate
$\sigma$ is relevant at the pure annihilation fixed point, that is, in
the beta function $\beta_\sigma=-y\sigma+O(\sigma^2)$, is $y>0$? If so, then as soon
as $\sigma\not=0$ it will flow away in the infrared into what is
presumably the fixed point controlling the active phase.
For $d>2$, the pure annihilation reaction is controlled by the gaussian
fixed point, so simple power counting suffices. This yields $y=2$,
consistent with the simulation results. For $d<2$, the annihilation 
fixed point is accessible within the $\epsilon=2-d$ expansion, and so
$y$ may be computed only perturbatively. The result is that\cite{Tauber}
\be\label{yeps}
y=2-\ffrac12m(m+1)\epsilon+O(\epsilon^2).
\ee
Notice that the $O(\epsilon)$ corrections are large in $d=1$, but no
conclusion may be drawn from this since the higher terms have been
neglected. 

Fortunately it is possible to compute $y$ exactly in $d=1$.\cite{Tauber}
There are
several ways of doing this, but the simplest is to realise that in this
limit it becomes a kind of a free fermion problem. This is because going
to the annihilation fixed point $g_R\to g^*$ corresponds to taking the
limit of the bare coupling $\lambda\to\infty$. In that case the term
$\lambda\ad^2a^2$ in (\ref{ham}) corresponds to an infinite hard core repulsion,
so that the particles behave in one dimension like free fermions, at
least in those periods of time evolution during which the other terms in
the hamiltonian do not play a role. (For this reason the problem is not
completely equivalent to free fermions.)
In that limit, it does not make sense to create the new particles at the
same lattice site. The best one can do is to distribute them between
$m$ neighbouring sites, so that the corresponding term in the lattice
hamiltonian has the form
\be
\sigma\sum_j\prod_{i=j-\frac m2}^{j+\frac m2}\cd_ic_j,
\ee
where $\cd_i$ and $c_j$ are now anticommuting operators. In the
continuum limit, we may make a Taylor expansion of each $\cd_i$ about
$i=j$, in powers of the lattice spacing $b$.
The lowest surviving term has the form
\be
\tilde\sigma\int\cd(\partial\cd)(\partial^2\cd)\ldots(\partial^m\cd)\,c\,dx,
\ee
where $\tilde\sigma=\sigma b^{m(m+1)/2}$ is now the effective expansion
parameter in the continuum limit. This extra factor modifies the
dimensional analysis, which then implies that
\be
y=2-\ffrac12m(m+1).
\ee
Thus the $O(\epsilon)$ result in (\ref{yeps}) appears to be exact for
$d=1$. We have no simple explanation of this, as we have also computed
explicitly the $O(\epsilon^2)$ terms and find them to be non-zero. 

However, this does imply that the branching is irrelevant at the
pure annihilation fixed point for $d=1$, and hence the infrared or late
time behaviour for sufficiently small $\sigma$ should be that of the
pure annihilation process, with finitely renormalised parameters (for
example, the diffusion constant.) It may also be shown that, even if the
original branching process does not allow for $m=2$ processes, these
will inevitably be generated under renormalisation, and, since this
coupling is the most relevant, it controls the late time behaviour.
Physically both these effects may be understood as
follows. In pure annihilation, the surviving particles become strongly
anticorrelated in space. This is because each sweeps out a region around
itself: for $d<2$, every test particle placed within that region has
probability one of eventually annihilating with it. (For $d>2$ the test
particle may escape.) When a small branching rate is turned on, the
single particles occasionally branch into bunches of $3,5,\ldots$
particles but these stay close together, and almost always annihilate
with their siblings before visiting other bunches. The effect is
therefore of diffusing bunches, which behave in many ways like single
particles with a reduced diffusion constant. Clearly even if branchings
only with $m>2$ are allowed, the pair annihilation process will generate
an effective $m=2$ branching rate.

For larger values of the branching rate $\sigma$ there should be a
transition to the active state, which should correspond to some
nontrivial fixed point of the \rg. But it seems to be very difficult to
analyse this within any perturbative \rg\ scheme. This is because the
problem has two critical dimensionalities: $d=2$ associated with the
nontrivial nature of the annihilation, and $d\approx\frac43$ where the
value of $y$ changes sign, and therefore no systematic
$\epsilon$-expansion is possible. So far we have not been able to find
another small parameter, and the best we can do is a truncated loop
expansion in fixed number of dimensions. This leads to the expected
fixed point, but the estimated values for the critical exponents are far
from those measured in simulations.
\subsection{Case of odd $m$} 
Although the above analysis might suggest that for $m=1$ the 
branching rate is relevant even when $d=1$, so that there is no
nontrivial transition, this is not the case, since now there is no
conservation law modulo 2, and the process $A\to0$ is immediately
generated under renormalisation. This has eigenvalue 2 and corresponds
to the generation of a mass gap in the theory. In fact one may show that
for small branching rates the mean density should decay exponentially to
zero. This conclusion is valid even when $d=2$: although in this case
the annihilation rate which generates the new term is irrelevant, it is
only logarithmically vanishing, and meanwhile the rate for the process
$A\to0$ is growing under the \rg. Once again, for sufficiently large
$\sigma$, there should be a transition to the active state. In this case
it is rather easier to analyse. On including the effective term
$\Delta(a-\ad a)$ in the hamiltonian, corresponding to $A\to0$, and
making the shift $\ad\to1+\ab$, one finds an interaction hamiltonian of
the form
\be
H_{int}=\int[\delta\ab a+\mu_1\ab a^2-\mu_2\ab^2a+\cdots]d^d\!x,
\ee
where $\mu_1$ and $\mu_2$ are positive constants, and $\delta$ may
change sign (as it does at the mean field transition.) The omitted terms
are of higher order, and their neglect is, this time, justified, since
there is no symmetry relating them to the lower order terms. This is a
well-known theory\cite{DP} which describes the universality class known as
directed percolation (DP), although it was first studied by particle
physicists in the context of 
reggeon field theory. Generically, any dynamical phase transition from
an inactive state, with no noise, to active one, is in the DP
universality class, and this has been verified for a number of models.
The branching and annihilating random walks for $m$ even and $d=1$ are
therefore an interesting exception to this general rule. They evade it
because they posses an additional conservation law. This is of course
quite a familiar idea from equilibrium critical behaviour.
\section{Conclusions}
These simple examples I hope illustrate the point that quantum field
theory still has many unexplored applications, which are not limited to
quantum systems nor to equilibrium critical behaviour. Perhaps we are
not yet at the stage when the mathematical beauty of such applications
is apparent, but the richness of the subject is such that I believe that
this may well emerge in the years to come.

\section*{Acknowledgements}

This work was carried out under partial support of EPSRC Grant GR/J78044.
I gratefully acknowledge my collaborators B.~P.~Lee, M.~Howard, and
U.~Tauber who have all contributed to the research described above.

\end{document}